\documentclass{amsart}
\usepackage{amsmath,amsthm,amsfonts,amssymb}

\theoremstyle{plain}
\newtheorem{lemme}{Lemma}[section]

\newtheorem{theoreme}[lemme]{Theorem}

\theoremstyle{definition}

\theoremstyle{remark}
\newtheorem*{remarque}{Remark}

\newcommand{\R}{\mathbb{R}}
\newcommand{\N}{\mathbb{N}}
\newcommand{\Z}{\mathbb{Z}}

\newcommand{\e}{\epsilon}
\newcommand{\w}{\omega}

\newcommand{\f}{\frac}
\newcommand{\h}{\hbar}

\begin{document}

\title{The Klein-Gordon's field. A counter-example of the 
classical limit}

\author{Jaume Haro}

\footnote{Partially supported by DGESIC (spain), project 
PB98-0932-C02-01.}

\date{26 Febraury, 2000}

\address{
Departament de Matem\`atica Aplicada I\\
E.T.S.E.I.B., Universitat Polit\`ecnica de Catalunya
}

\email{haro@ma1.upc.es}

\maketitle

\begin{minipage}[]{4.5in}
{\bf Abstract}.    {\scriptsize      {\sl
We will study the Klein-Gordon's field with an homogeneous external
 potential,
which does not depend on $\h$. We will construct the Fock's space
corresponding to our problem and we will see that there are phenomena
of creation and anihilation of pairs particle-antiparticle. Finally,
we will see that in dimension $1$, when $\h\rightarrow 0$, 
these phenomena
disappear. However, in dimension $2$ or $3$, when $\h\rightarrow 0$, the
creation probability of particle-antiparticle pairs is not zero.
 }}\end{minipage}

\section{Introduction}
In this work we will study the classical limit of Klein-Gordon's
field, with an homogeneous potential which does not depend on Planck's
constant.

First we will see that, in this case, the Klein-Gordon's equation is
equivalent to a hamiltonian system, composed by an infinite number of
harmonic oscillators with frequencies which depend on time. Once we
have seen this equivalence, we will quantize these oscillators and we
will obtain the energy and the electric charge operators. With the
energy operator, we will obtain the quantum equation of Klein-Gordon's
field. We will also see that we can find all the eigenfunctions of the
energy and the electric charge operators. Consequently, with all those
eigenfunctions we can construct the Fock's space.

After that, we will study the quantum dynamic of vacuum state. We will
see that, if the space dimension is $1$, when $\h\rightarrow 0$, the
 probability that does not  exist any 
particle-antiparticle pair,
converges to $1$.
However, in dimension $2$ or $3$, we will prove that, when
$\h\rightarrow 0$, this probability does not converge to $1$. 
Consequently, in dimension $2$ or $3$,  the classical
limit is not true.

The notation that we are going to use, is the following

$<,>$  euclidean scalar product. 

$<,>_2$ scalar product of ${\mathcal L}^2$. 

$||.||_2$ norm ${\mathcal L}^2$.

$||.||_2$ norm $\infty$.

\section{The Klein-Gordon's field with an homogeneous potential}

To simplify, we will take $m=c=e=1$.

If we apply the Correspondence Principle $E\rightarrow
i\hbar\partial_t$, $\vec p\rightarrow
-i\hbar\vec\nabla$ to the
 relativistic relation
$E^2=|\vec p+\vec f(t)|^2+1$, we obtain the Klein-Gordon's equation, 

$$-\h^2\partial_t^2\psi=|-i\hbar\vec\nabla+\vec f(t)|^2\psi+\psi.$$

One important property of the K-G's equation is the electric charge
conservation $\dot{\rho}(t)=0$, where
 $\rho(t)=<i\h\partial_t\psi,\psi>_2+
<\psi,i\h\partial_t\psi>_2$. However there is no norme square
conservation  
$\dot{||\psi(t)||}_2^2\not=0$.
Then, to be able to speak about probabilities, we have to consider that
the  K-G's field describes an infinity of harmonic oscillators.
After that, we have to quantize these oscillators to arrive to an
equation of this type,
$i\hbar\partial_t|\Phi>=H|\Phi>$,
where $H$ is an self-adjoint operator.

\subsection{ The Quantization of Klein-Gordon's field}

Suppose that the domain is finite. To simplify we take the
 n-dimensional interval  $[-\pi,\pi]^n$.

The lagrangian and the energy of the system at $t$ time are:

$$L(t)=\h^2||\partial_t\psi||_2^2-
||(-i\hbar\vec\nabla+\vec f(t))\psi||_2^2-||\psi||_2^2$$

$$E(t)=\h^2||\partial_t\psi||_2^2+
||(-i\hbar\vec\nabla+\vec f(t))\psi||_2^2+||\psi||_2^2.$$

We expand $\psi$ in Fourier's serie,
$\psi(\vec x,t)=\sum_{\vec k\in\Z^n}A_{\vec k}(t)\psi_{\vec k}(\vec
x)$, with $\psi_{\vec k}(\vec x)=\f{e^{i<\vec k,\vec
x>}}{(2\pi)^{\f{n}{2}}}$. Then

$$L(t)=\sum_{\vec k\in\Z^n}\h^2|\dot{A}_{\vec k}|^2- 
\e_{\vec k}^2(t)|A_{\vec k}|^2, \quad
 \mbox{ where } \e_{\vec k}(t)=\sqrt{|\h\vec k+\vec f(t)|^2+1}.$$

With the  momenta  $B_{\vec k}=\h^2 \dot{A}_{\vec k}$,
we obtain
 
$$E(t)=\sum_{\vec k\in\Z^n}\f{|B_{\vec k}|^2}{\h^2}+ 
\e_{\vec k}^2(t)|A_{\vec k}|^2, \quad
\rho(t)=\sum_{\vec k\in\Z^n}\f{i}{\h}(A^*_{\vec k}B_{\vec k}-
A_{\vec k}B^*_{\vec k}).$$

We make the real canonical change
$$B_{\vec k}=\f{\h}{\sqrt{2}}(P_{\vec k}+i\bar{P}_{\vec k});\quad
A_{\vec k}=\f{1}{\h\sqrt{2}}(Q_{\vec k}+i\bar{Q}_{\vec k}),$$
and let $\w_{\vec k}(t)=\f{\e_{\vec k}(t)}{\h}$ be the corresponding 
frequency, then $E(t)$ and
$\rho(t)$ take the form

$$E(t)=\f{1}{2}\sum_{\vec k\in\Z^n}(P_{\vec k}^2+
\w_{\vec k}^2(t)Q_{\vec k}^2)+(\bar{P}_{\vec k}^2+
\w_{\vec k}^2(t)\bar{Q}_{\vec k}^2)$$
$$\rho(t)=\f{1}{\h}\sum_{\vec k\in\Z^n}(\bar{Q}_{\vec k}P_{\vec k}-
Q_{\vec k}\bar{P}_{\vec k}).$$

That is the energy decomposition in oscillators. Notice, the
K-G's equation is equivalent to the hamiltonian system

\begin{eqnarray*}\left\{\begin{array}{ccc}
\dot{Q}_{\vec k} &=& P_{\vec k} \\
\dot{P}_{\vec k} &=&-\w_{\vec k}^2(t) Q_{\vec k}
\end{array}\right. \quad
\left\{\begin{array}{ccc}
\dot{\bar{Q}}_{\vec k} &=& \bar{P}_{\vec k} \\
\dot{\bar{P}}_{\vec k} &=&-\w_{\vec k}^2(t) \bar{Q}_{\vec k}
\end{array}\right. \end{eqnarray*}

Now, to obtain the quantum theory, what we have to do is to quantize
these oscillators, i.e.
$P_{\vec k}\rightarrow -i\h\partial_{Q_{\vec k}}$, 
$\bar{P}_{\vec k}\rightarrow -i\h\partial_{\bar{Q}_{\vec k}}$, and the
equation will be

$$i\h\partial_t\Phi=
\f{1}{2}\sum_{\vec k\in\Z^n}[(-\h^2\partial^2_{Q_{\vec k}}+
\w_{\vec k}^2(t)Q_{\vec k}^2)+(-\h^2\partial^2_{\bar{Q}_{\vec k}}+
\w_{\vec k}^2(t)\bar{Q}_{\vec k}^2)]\Phi-\sum_{\vec k\in\Z^n}
\w_{\vec k}(t)\Phi.$$

Now, we will look for the eigenfunctions of the energy and of the 
electric
charge operators. First, we have to introduce the creation and
anihilation operators for particles and antiparticles

$$a_{\vec k}(t)=\f{1}{2\sqrt{\e_{\vec k}(t)}}[(\h\partial_{Q_{\vec k}}
+\w_{\vec k}(t)Q_{\vec k})+i(\h\partial_{\bar{Q}_{\vec k}}
+\w_{\vec k}(t)\bar{Q}_{\vec k})]$$
$$a^+_{\vec k}(t)=\f{1}{2\sqrt{\e_{\vec k}(t)}}[(-\h\partial_{Q_{\vec k}}
+\w_{\vec k}(t)Q_{\vec k})-i(-\h\partial_{\bar{Q}_{\vec k}}
+\w_{\vec k}(t)\bar{Q}_{\vec k})]$$
$$b_{-\vec k}(t)=\f{1}{2\sqrt{\e_{\vec k}(t)}}[(\h\partial_{Q_{\vec k}}
+\w_{\vec k}(t)Q_{\vec k})-i(\h\partial_{\bar{Q}_{\vec k}}
+\w_{\vec k}(t)\bar{Q}_{\vec k})]$$
$$b^+_{-\vec k}(t)=\f{1}{2\sqrt{\e_{\vec k}(t)}}
[(-\h\partial_{Q_{\vec k}}
+\w_{\vec k}(t)Q_{\vec k})+i(-\h\partial_{\bar{Q}_{\vec k}}
+\w_{\vec k}(t)\bar{Q}_{\vec k})].$$

Then
$$E(t)=\sum_{\vec k\in\Z^n}\e_{\vec k}(t)(a^+_{\vec k}(t)
a_{\vec k}(t)+b^+_{-\vec k}(t)b_{-\vec k}(t))$$

$$\rho(t)=\sum_{\vec k\in\Z^n}(a^+_{\vec k}(t)
a_{\vec k}(t)-b^+_{-\vec k}(t)b_{-\vec k}(t)).$$

We construct {\bf the vacuum state at  $t$ time}.

Let's consider $\phi_{\vec k}^{0,0}(Q_{\vec k},\bar{Q}_{\vec k},t)=
\sqrt{\f{\w_{\vec k}(t)}{\pi\h}}e^{-\f{\w_{\vec k}(t)}{2\h}(
Q^2_{\vec k}+\bar{Q}^2_{\vec k})}$, then the vacumm state at
$t$ time, $|0>(t)$, is

$$|0>(t)=\prod_{\vec k\in\Z^n}\phi_{\vec k}^{0,0}(Q_{\vec
k},\bar{Q}_{\vec k},t),$$
because

$$E(t)|0>(t)=0
\quad \rho(t)|0>(t)=0.$$

Starting from this state we will define all the  others. In fact, 

the state $|1^+_{\vec k}>(t)=a^+_{\vec k}(t)|0>(t)$, verifies

$$E(t)|1^+_{\vec k}>(t)=\e_{\vec k}(t)|1^+_{\vec k}>(t)
\quad \rho(t)|1^+_{\vec k}>(t)=|1^+_{\vec k}>(t),$$
consequently, $|1^+_{\vec k}>(t)$ is the state of a particle with
energy $\e_{\vec k}(t)$ at $t$ time.

The state $|1^-_{\vec k}>(t)=b^+_{\vec k}(t)|0>(t)$, verifies

$$E(t)|1^-_{\vec k}>(t)=\e_{-\vec k}(t)|1^-_{\vec k}>(t)
\quad \rho(t)|1^-_{\vec k}>(t)=-|1^-_{\vec k}>(t),$$
consequently, $|1^-_{\vec k}>(t)$ is the state of an antiparticle with
energy $\e_{-\vec k}(t)$ at  $t$ time.

In general, we consider  series
\begin{eqnarray*}
\{n_{\vec k}\}:\left. \begin{array}{ccc}
\Z^n &\rightarrow &\N\\
\vec k &\rightarrow  & n_{\vec k}
\end{array}\right.\end{eqnarray*}
and let
$$|\{n_{\vec k}\};\{m_{\vec k}\}>(t)=
\prod_{\vec k\in\Z^n}\f{(a^+_{\vec k}(t))^{n_{\vec k}}}{\sqrt{n_{\vec
k}!}}\f{(b^+_{-\vec k}(t))^{m_{\vec k}}}{\sqrt{m_{\vec
k}!}}|0>(t).$$

Then $|\{n_{\vec k}\};\{m_{\vec k}\}>(t)$, verifies

$$E(t)|\{n_{\vec k}\};\{m_{\vec k}\}>(t)=\sum_{\vec l\in\Z^n}
\e_{\vec l}(t)(n_{\vec l}+m_{\vec l})
|\{n_{\vec k}\};\{m_{\vec k}\}>(t)$$

$$\rho(t)|\{n_{\vec k}\};\{m_{\vec k}\}>(t)=\sum_{\vec l\in\Z^n}
(n_{\vec l}-m_{\vec l})
|\{n_{\vec k}\};\{m_{\vec k}\}>(t).$$

Consequently, the state
 $|\{n_{\vec k}\};\{m_{\vec k}\}>(t)$ contains,
at $t$ time,  $n_{\vec k}$ particles 
 and  $m_{\vec k}$ antiparticles with energy
$\e_{\vec k}(t)$, for each $\vec k\in\Z^n$. 

\section{The counter-example}
\subsection{Quantum dynamic}
First, we study the case
 $\vec f(t)\equiv \vec 0$, then

$$E=\f{1}{2}\left[\sum_{\vec k\in\Z^n}(-\h^2\partial^2_{Q_{\vec k}}+
\w_{\vec k}^2Q_{\vec k}^2)+(-\h^2\partial^2_{\bar{Q}_{\vec k}}+
\w_{\vec k}^2\bar{Q}_{\vec k}^2)\right]-
\sum_{\vec k\in\Z^n}\w_{\vec k},$$
where $\w_{\vec k}=\f{\sqrt{|\h\vec k|^2+1}}{\h}$.
Notice that the energy does not depend on time, then the eigenvalues
 $|\{n_{\vec k}\};\{m_{\vec
k}\}>(t)\equiv |\{n_{\vec k}\};\{m_{\vec k}\}>$ do
not depend on time. Therefore, the solution of the problem 

\begin{eqnarray*}\left\{\begin{array}{ccc}
i\h\partial_t|\Psi>&=&E|\Psi>\\
|\Psi>(0)&=&|\{n_{\vec k}\};\{m_{\vec k}\}>,
\end{array}\right.\end{eqnarray*}
is

$$T^t_q|\{n_{\vec k}\};\{m_{\vec k}\}>=
e^{-\f{i}{\h}\sum_{\vec l\in\Z^n}
\e_{\vec l}(n_{\vec l}+m_{\vec l})t}|\{n_{\vec k}\};\{m_{\vec k}\}>.$$

In particular, $T^t_q|0>=|0>$, i.e., when $\vec f(t)\equiv 0$,
the vacuum state is invariant for the quantum dynamic, and there is no
 creation and anihilation particle-antiparticle pairs.

We now study the vacuum dynamic when
$\vec f(t)\not= \vec 0$. Let $T^t_q|0>(0)$ be the solution of the problem

\begin{eqnarray*}\left\{\begin{array}{ccc}
i\h\partial_t|\Psi>&=&E(t)|\Psi>\\
|\Psi>(0)&=&|0>(0),
\end{array}\right.\end{eqnarray*}
then $T^t_q|0>(0)=\prod_{\vec k\in\Z^n}T^t_{\h}
\phi_{\vec k}^{0,0}(Q_{\vec k},\bar{Q}_{\vec k},0)$, where
$T^t_{\h}\phi_{\vec k}^{0,0}(Q_{\vec k},\bar{Q}_{\vec k},0)$ 
is the solution of problem 

\begin{eqnarray}\label{A}\left\{\begin{array}{ccc}
i\h\partial_t\phi &=&
[\f{1}{2}(-\h^2\partial^2_{Q_{\vec k}}+
\w_{\vec k}^2(t)Q_{\vec k}^2
-\h^2\partial^2_{\bar{Q}_{\vec k}}+
\w_{\vec k}^2(t)\bar{Q}_{\vec k}^2)-\w_{\vec k}(t)]\phi\\
\phi(0) &=&
\phi_{\vec k}^{0,0}(Q_{\vec k},\bar{Q}_{\vec k},0).
\end{array}\right.\end{eqnarray}

Denote by $P^0_{\h}(t)=|(t)<0|T^t_q|0>(0)|^2$,  the probability that 
it does not exist any particle-antiparticle pair at $t$ time.

Then, we have the

\begin{theoreme} \label{T}

Let $n$ be the dimension of the space and suppose that $\vec f\in
{\mathcal C}_0^{\infty}(0,T)$, then:

If $n=1$ we have

$$\lim_{\h\rightarrow 0}P^0_{\h}(t)=1 \quad \forall t\in \R.$$

If $n=2$ or $3$, at $t$ time such that 
$\dot{\vec{f}}(t)\not= \vec 0$, we have 
$$\lim_{\h\rightarrow 0}P^0_{\h}(t)\not=1 .$$

\end{theoreme}

Consequently, in the case $n=2$ or $3$, at  $t$ time such that
 $\dot{\vec f}(t)\not= \vec 0$, we do not obtain the classical limit.

\begin{remarque} In dimension $1$ the result is valid for
periodic potentials, i.e., for $f(x,t)=\sum_{k=0}^N[
f_k(t)\sin(kx)+g_k(t)\cos(kx)]$, we have
$\lim_{\h\rightarrow 0}P^0_{\h}(t)=1$.
\end{remarque}

\begin{remarque} Le Theorem \ref{T} is valid for the 
Dirac's field. 
\end{remarque}

\section{Proofs}

To make the proof of theorem we need the following

\begin{lemme} \label{L}

The solution of the problem (\ref{A}) is

$$T^t_{\h}\phi^{0,0}_{\vec k}(0)=A_{\vec k}(t)\phi^{0,0}_{\vec k}(t)
+\left(-\f{i\h\dot{\e}_{\vec k}(t)}{4\e_{\vec k}^2(t)}
+\h^2 B_{\vec k}(t)\right)\phi^{1,1}_{\vec k}(t)
+\h^2 \gamma_{\vec k}(t),$$
with

 $|1-|A_{\vec k}(t)|^2|\leq \f{\h^2 K}{\e_{\vec k}^4}$.

$|B_{\vec k}(t)|^2, ||\gamma_{\vec k}(t)||_2^2\leq \f{K}{\e_{\vec k}^4}$;
$\gamma_{\vec k}(t)\bot\phi^{0,0}_{\vec k}(t),\phi^{1,1}_{\vec k}(t)$.

Where, 

$\phi^{1,1}_{\vec k}(t)=a^+_{\vec k}(t)b^+_{-\vec k}(t)
\phi^{0,0}_{\vec k}(t)$.

$K$ is a constant independent on $\vec k$, $\h$ and $t$.

$\e_{\vec k}=\sqrt{|\h\vec k|^2+1}$.

\end{lemme}

With this lemma we can make the  

{\bf Proof  of Theorem \ref{T}:}

If $n=1$,  $P^0_{\h}(t)=\prod_{k\in\Z}
|A_k(t)|^2$. We write, $|A_k(t)|^2=1+\bar{A}_k(t)$, then

\begin{eqnarray*}
P^0_{\h}(t)=1+\f{1}{1!}\sum_{k\in\Z}\bar{A}_k(t)+
\f{1}{2!}\sum_{\substack{k_1,k_2\in\Z \\ k_1\not= k_2}}\bar{A}_{k_1}(t)
\bar{A}_{k_2}(t)+
\f{1}{3!}\sum_{\substack{k_1,k_2,k_3\in\Z \\ k_i\not=
k_j;i\not=j \\i,j=1,2,3 }}
\bar{A}_{ k_1}(t)\bar{A}_{ k_2}(t)\bar{A}_{ k_3}(t)
+\cdots
\end{eqnarray*}

We bound

\begin{eqnarray*}&&
|P^0_{\h}(t)-1|\leq
\f{1}{1!}\sum_{k\in\Z}|\bar{A}_{k}(t)|+
\f{1}{2!}
\sum_{\substack{k_1,k_2\in\Z \\ k_1\not= k_2}}|\bar{A}_{ k_1}(t)|
|\bar{A}_{k_2}(t)|+\cdots\leq
\sum_{n=1}^{\infty}\f{1}{n!}\left(\sum_{k\in\Z}|\bar{A}_{ k}(t)|\right)^n
\\ &&\leq \sum_{n=1}^{\infty}\f{1}{n!}\left(K\h\sum_{k\in\Z} 
\f{\h}{\e_{k}^4}\right)^n
\leq \sum_{n=1}^{\infty}\f{1}{n!}\left(K\h\sum_{k\in\Z} 
\f{\h}{\e_{k}^2}\right)^n,
\end{eqnarray*}
since $\sum_{k\in\Z} \f{\h}{\e_{ k}^2}\leq
\int_{\R}\f{dx}{x^2+1}+\h=\pi+\h$, we have

\begin{eqnarray*}
|P^0_{\h}(t)-1|\leq \sum_{n=1}^{\infty}\f{1}{n!}(K\h(\pi+\h))^n=
e^{K\h(\pi+\h)}-1,
\end{eqnarray*}
therefore
$$\lim_{\h\rightarrow 0}P^0_{\h}(t)=1.$$

We now study the case $n=2$. The case  $n=3$ is analogous.

Denote by $P^1_{\h}(t)=\sum_{\vec k\in\Z^2}|(t)<1^+_{\vec k}1^-_{-\vec k}
|T^t_q|0>(0)|^2$, the probability that at $t$ time, does exist a
 particle-antiparticle pair, then

$$P^1_{\h}(t)=\sum_{\vec k\in\Z^2}\left|
-\f{i\h\dot{\e}_{\vec k}(t)}{4\e_{\vec k}^2(t)}
+\h^2 B_{\vec k}(t)\right|^2
\prod_{\substack{\vec l \in \Z^2\\ \vec l\not=\vec k}}
|A_{\vec l}(t)|^2.$$

We calcule 

\begin{eqnarray*}&&
\lim_{\h\rightarrow 0}\left|P^1_{\h}(t)-
\sum_{\vec k\in \Z^2}
\f{\h^2}{16}\f{\dot{\e}^2_{\vec k}(t)}{\e^4_{\vec k}(t)}
P^0_{\h}(t)\right|\leq
\lim_{\h\rightarrow 0}\left|P^1_{\h}(t)-
\sum_{\vec k\in \Z^2}
\f{\h^2}{16}\f{\dot{\e}^2_{\vec k}(t)}{\e^4_{\vec k}(t)}
\prod_{\substack{\vec l \in \Z^2\\ \vec l\not=\vec k}}|A_{\vec
l}(t)|^2\right |
\\ &&+\lim_{\h\rightarrow 0}\left|
\sum_{\vec k\in \Z^2}
\f{\h^2}{16}\f{\dot{\e}^2_{\vec k}(t)}{\e^4_{\vec k}(t)}
(1-|A_{\vec k}(t)|^2)
\prod_{\substack{\vec l \in \Z^2\\ \vec l\not=\vec k}}|A_{\vec
l}(t)|^2\right |
\\ &&
\leq \lim_{\h\rightarrow 0}\sum_{\vec k\in \Z^2}\left(
\f{\h^3}{2}\f{|\dot{\e}_{\vec k}(t)|}{\e^2_{\vec k}(t)}|B_{\vec k}(t)|
+\h^4 |B_{\vec k}(t)|^2+\f{\h^2}{16}
\f{\dot{\e}^2_{\vec k}(t)}{\e^4_{\vec k}(t)}
|1-|A_{\vec k}(t)|^2|\right).
\end{eqnarray*}

Because of the lemma  (\ref{L}) and the relation 
$\e^2_{\vec k}\leq C\e^2_{\vec k}(t)$ where $C=2(1+||{\vec
f}||^2_{\infty})$, we obtain

\begin{eqnarray*}
&&\lim_{\h\rightarrow 0}\left|P^1_{\h}(t)-
\sum_{\vec k\in \Z^2}
\f{\h^2}{16}\f{\dot{\e}^2_{\vec k}(t)}{\e^4_{\vec k}(t)}
P_{\h}(t)\right|\\&&\leq
\lim_{\h\rightarrow 0}
\sum_{\vec k\in \Z^2}\f{\h^2}{\e^4_{\vec k}}
\lim_{\h\rightarrow
0}\h\left(\sqrt{K}C||\dot{\vec{f}}||_{\infty} 
+K\h(||\dot{\vec{f}}||^2_{\infty}+1) \right)=0,
\end{eqnarray*}

because $\lim_{\h\rightarrow 0}
\sum_{\vec k\in \Z^2}\f{\h^2}{\e^4_{\vec k}}=\int_{\R^2}
\f{d{\vec x}^2}{(|{\vec x}|^2+1)^2}=\pi$.

Therefore, we have proved
that

$$\lim_{\h\rightarrow 0}P^1_{\h}(t)=
\lim_{\h\rightarrow 0}
\sum_{\vec k\in \Z^2}
\f{\h^2}{16}\f{\dot{\e}^2_{\vec k}(t)}{\e^4_{\vec k}(t)}P^0_{\h}(t).$$

With this result, we can prove that for
$n=2$, if  $\dot{\vec f}(t)\not= \vec 0$, then
 $\lim_{\h\rightarrow 0}P^0_{\h}(t)\not= 1$. In fact, we take
$t_0$ such that $\dot{\vec f}(t_0)\not= \vec 0$ and we assume that
$\lim_{\h\rightarrow 0}P^0_{\h}(t_0)= 1$.
Thus, $\lim_{\h\rightarrow 0}P^1_{\h}(t_0)=0$.

However
\begin{eqnarray*}&&
\lim_{\h\rightarrow 0}P^1_{\h}(t_0)=
\lim_{\h\rightarrow 0}
\sum_{\vec k\in \Z^2}
\f{\h^2}{16}\f{\dot{\e}^2_{\vec k}(t_0)}{\e^4_{\vec k}(t_0)}
\lim_{\h\rightarrow 0}P^0_{\h}(t_0)
=(\mbox{for hypothesis})=\\&&
\lim_{\h\rightarrow 0}
\sum_{\vec k\in \Z^2}
\f{\h^2}{16}\f{\dot{\e}^2_{\vec k}(t_0)}{\e^4_{\vec k}(t_0)}
=\f{1}{16}\int_{\R^2}
\f{<\dot{\vec f}(t_0),\vec x>^2}
{(|\vec x|^2+1)^3}d\vec x^2\not= 0,\end{eqnarray*}
because $\dot{\vec f}(t_0)\not= \vec 0$. Therefore, we have a 
contradiction and in
consequence, $\lim_{\h\rightarrow
0}P^0_{\h}(t_0)\not= 1$.
\qed

Now, we make the 

{\bf Proof of lemma \ref{L}:}

First, we will construct a semi-classical solution of the problem 
(\ref{A}). To search a semi-classical solution,
 we have to consider the functions
$\phi_{\vec k}^{s,s}(t)=\f{(a_{\vec k}^+(t))^s
(b_{-\vec k}^+(t))^s}{s!}\phi_{\vec k}^{0,0}(t)$ with $s\in\N$.

We write the problem (\ref{A}) in the  form
\begin{eqnarray*}\left\{\begin{array}{ccc}
i\h\partial_t\phi&=&H_{\vec k}(t)\phi\\
\phi(0)&=&\phi_{\vec k}^{0,0}(0),\end{array}\right.\end{eqnarray*}
where $H_{\vec k}(t)=\e_{\vec k}(t)(a_{\vec k}^+(t)a_{\vec k}(t)+
b_{-\vec k}^+(t)b_{-\vec k}(t))$. We expand the solution in powers
serie of $\h$, in the following form,
$T^t_{\h}\phi_{\vec k}^{0,0}(0)=\sum_{j,s\in\N}\h^{s+j}
A_{s,\vec k}^j(t)\phi_{\vec k}^{s,s}(t)$. Then, because of following

\begin{lemme}
$$\dot{\phi}_{\vec k}^{s,s}(t)=\f{\dot{\e}_{\vec k}(t)}
{2{\e}_{\vec k}(t)}(s{\phi}_{\vec k}^{s-1,s-1}(t)-(s+1)
{\phi}_{\vec k}^{s+1,s+1}(t)),$$
\end{lemme}
we obtain, after having equalized the powers of $\h$, the
 system:

If $s=0$
$$\dot{A}_{0,\vec k}^0=0;\quad
\dot{A}_{0,\vec k}^j+\f{\dot{\e}_{\vec k}(t)}
{2{\e}_{\vec k}(t)}A_{1,\vec k}^{j-1}=0, \mbox{ for } j>0.$$

If $s>0$

$$-i\f{\dot{\e}_{\vec k}(t)}{2{\e}_{\vec k}(t)}A_{s-1,\vec k}^0
-2{\e}_{\vec k}(t)A_{s,\vec k}^0=0.$$
$$i\dot{A}_{s,\vec k}^0-
i\f{\dot{\e}_{\vec k}(t)}{2{\e}_{\vec k}(t)}sA_{s-1,\vec k}^1
-2s{\e}_{\vec k}(t)A_{s,\vec k}^1=0.$$
$$i\dot{A}_{s,\vec k}^{j-1}+
i\f{\dot{\e}_{\vec k}(t)}{2{\e}_{\vec k}(t)}
\left((s+1)A_{s+1,\vec k}^{j-2}-sA_{s-1,\vec k}^{j}\right)
-2s{\e}_{\vec k}(t)A_{s,\vec k}^j=0,\mbox{ for } j>1.$$

We obtain the solution of the system by recurrence. In fact 

$$A_{0,\vec k}^{0}(t)\equiv 1;\quad A_{1,\vec k}^{0}(t)=
-i\f{\dot{\e}_{\vec k}(t)}{4{\e}_{\vec k}^2(t)};
\quad A_{0,\vec k}^{1}(t)=
\int_0^ti\f{\dot{\e}^2_{\vec k}(\tau)}{8{\e}_{\vec k}^3(\tau)}d\tau.$$
$$A_{1,\vec k}^{1}(t)=\f{1}{2{\e}_{\vec k}(t)}
(i\dot{A}_{1,\vec k}^0-
i\f{\dot{\e}_{\vec k}(t)}{2{\e}_{\vec k}(t)}A_{0,\vec k}^1);\quad
A_{2,\vec k}^{0}(t)=
-i\f{\dot{\e}_{\vec k}(t)}{4{\e}_{\vec k}^2(t)}
A_{1,\vec k}^{0}(t).$$
$$A_{0,\vec k}^{2}(t)=-
\int_0^t\f{\dot{\e}_{\vec k}(\tau)}{2{\e}_{\vec k}(\tau)}
A_{1,\vec k}^{1}(\tau)d\tau;
A_{1,\vec k}^{2}(t)=\f{1}{2{\e}_{\vec k}(t)}
(i\dot{A}_{1,\vec k}^1+
i\f{\dot{\e}_{\vec k}(t)}{2{\e}_{\vec k}(t)}
(2A_{2,\vec k}^0-A_{0,\vec k}^2))$$
$$A_{3,\vec k}^{0}(t)=
-i\f{\dot{\e}_{\vec k}(t)}{4{\e}_{\vec k}^2(t)}
A_{2,\vec k}^{0}(t);\quad
A_{0,\vec k}^{3}(t)=-
\int_0^t\f{\dot{\e}_{\vec k}(\tau)}{2{\e}_{\vec k}(\tau)}
A_{1,\vec k}^{2}(\tau)d\tau$$
$$
A_{2,\vec k}^{1}(t)=\f{1}{4{\e}_{\vec k}(t)}
(i\dot{A}_{2,\vec k}^0-
i\f{\dot{\e}_{\vec k}(t)}{{\e}_{\vec k}(t)}A_{1,\vec k}^1);\quad etc.$$

With these solutions, and the relation
${\e}^2_{\vec k}\leq C {\e}^2_{\vec k}(t)$, we obtain the 

\begin{lemme}\label{le}
If $s,j\leq 3$ we have
$$|A_{s,\vec k}^{j}(t)|\leq \f{\bar{C}}{\e^{2s+j}_{\vec k}}
\mbox{ for } s>0; \quad 
|A_{0,\vec k}^{j}(t)|\leq \f{\bar{C}}{\e^{2+j}_{\vec k}}
\mbox{ for } j>0$$

$$|\dot{A}_{s,\vec k}^{j}(t)|\leq \f{g(t)}{\e^{2s+j}_{\vec k}}
\mbox{ for } s>0; \quad 
|\dot{A}_{0,\vec k}^{j}(t)|\leq \f{g(t)}{\e^{2+j}_{\vec k}}
\mbox{ for } j>0,$$
where $\bar{C}$ is a constant independent on $\vec k$, and
$g(t)\in {\mathcal C}_0^{\infty}(0,T)$ is a function independent on
$\vec k$.
\end{lemme}

Now, we show that the function

\begin{eqnarray*}
&&\bar{\phi}_{\vec k}(t)=(A_{0,\vec k}^{0}+\h A_{0,\vec k}^{1}+
\h^2 A_{0,\vec k}^{2}+\h^3 A_{0,\vec k}^{3})\phi_{\vec k}^{0,0}(t)
+(\h A_{1,\vec k}^{0}+
\h^2 A_{1,\vec k}^{1}\\&&+\h^3 A_{1,\vec k}^{2})\phi_{\vec k}^{1,1}(t)
+(\h^2 A_{2,\vec k}^{0}+\h^3 A_{2,\vec k}^{1})\phi_{\vec k}^{2,2}(t)
+\h^3 A_{3,\vec k}^{0}\phi_{\vec k}^{3,3}(t)\end{eqnarray*}
is a semi-classical solution. In fact, we calcule

\begin{eqnarray*}
(i\h\partial_t-H_{\vec k})\bar{\phi}_{\vec k}(t) &=&
-2i\h^4\f{\dot{\e}_{\vec k}(t)}{{\e}_{\vec k}(t)}A_{3,\vec k}^0
\phi_{\vec k}^{4,4}(t)+\h^4(i\dot{A}_{3,\vec k}^0-
i\f{3\dot{\e}_{\vec k}(t)}{2{\e}_{\vec k}(t)}A_{2,\vec k}^1)
\phi_{\vec k}^{3,3}(t)\\&&+\h^4(i\dot{A}_{2,\vec k}^1+
i\f{\dot{\e}_{\vec k}(t)}{2{\e}_{\vec k}(t)}
(3A_{3,\vec k}^0-2A_{1,\vec k}^2))\phi_{\vec k}^{2,2}(t)\\&&+
\h^4(i\dot{A}_{1,\vec k}^2+
i\f{\dot{\e}_{\vec k}(t)}{2{\e}_{\vec k}(t)}
(2A_{2,\vec k}^1-A_{0,\vec k}^3))\phi_{\vec k}^{1,1}(t).
\end{eqnarray*}

We deduce from the lemma (\ref{le}), that 

$$||(i\h\partial_t-H_{\vec k})\bar{\phi}_{\vec k}(t)||_2^2\leq
\f{2\h^8}{{\e}^8_{\vec k}}(3g^2(t)+14C\bar{C}^2|\dot{\vec f}(t)|^2).$$

Furthermore, if using that

$$||T^t_{\h}\phi_{\vec k}^{0,0}(0)-\bar{\phi}_{\vec k}(t)||_2
\leq\f{1}{\h}\int_0^t
||(i\h\partial_{\tau}-H_{\vec k}(\tau))\bar{\phi}_{\vec k}(\tau)||_2
d\tau,$$
we obtain

\begin{eqnarray*}
&&||T^t_{\h}\phi_{\vec k}^{0,0}(0)-\bar{\phi}_{\vec k}(t)||_2
\leq \f{\sqrt{2}\h^3}{{\e}^4_{\vec k}}\int_0^t
\sqrt{3g^2(\tau)+14C\bar{C}^2|\dot{\vec f}(\tau)|^2}d\tau\\&&\leq
\f{\sqrt{2}\h^3}{{\e}^4_{\vec k}}\int_0^T
\sqrt{3g^2(\tau)+14C\bar{C}^2|\dot{\vec f}(\tau)|^2}d\tau
\equiv\f{\h^3\tilde{C}}{{\e}^4_{\vec k}}.\end{eqnarray*}

Therefore, $T^t_{\h}\phi_{\vec k}^{0,0}(0)$ has the form

\begin{eqnarray*}
&&T^t_{\h}\phi_{\vec k}^{0,0}(0)=(A_{0,\vec k}^{0}+\h A_{0,\vec k}^{1}+
\h^2 A_{0,\vec k}^{2}+\h^3 A_{0,\vec k}^{3}+\h^3 F_{\vec k})
\phi_{\vec k}^{0,0}(t)
+(\h A_{1,\vec k}^{0}+
\h^2 A_{1,\vec k}^{1}\\&&+\h^3 A_{1,\vec k}^{2}+\h^3 G_{\vec k})
\phi_{\vec k}^{1,1}(t)
+(\h^2 A_{2,\vec k}^{0}+\h^3 A_{2,\vec k}^{1}+\h^3 I_{\vec k})
\phi_{\vec k}^{2,2}(t)+\h^3 \beta_{\vec k}(t),
\end{eqnarray*}
with $|F_{\vec k}(t)|,|G_{\vec k}(t)|,|I_{\vec k}(t)|\leq
\f{\tilde{C}}{{\e}^4_{\vec k}}$,  $||\beta_{\vec k}(t)||_2
\leq \f{\tilde{C}+\bar{C}}{{\e}^4_{\vec k}}$
and $\beta_{\vec k}(t)\bot \phi_{\vec k}^{0,0}(t),
\phi_{\vec k}^{1,1}(t), \phi_{\vec k}^{2,2}(t).$

Finally, if we take
$$A_{\vec k}(t)=A_{0,\vec k}^{0}(t)+\h A_{0,\vec k}^{1}(t)+
\h^2 A_{0,\vec k}^{2}(t)+\h^3 A_{0,\vec k}^{3}(t)+\h^3 F_{\vec k}(t)$$
$$B_{\vec k}(t)=A_{1,\vec k}^{1}(t)+\h A_{1,\vec k}^{2}(t)+
\h G_{\vec k}(t)$$
$$\gamma_{\vec k}(t)=
( A_{2,\vec k}^{0}(t)+\h A_{2,\vec k}^{1}(t)+\h I_{\vec k}(t))
\phi_{\vec k}^{2,2}(t)+\h \beta_{\vec k}(t),$$
and $K=4(1+\bar{C}+\tilde{C})^2$, we obtain the proof of
lemma \ref{L}.
\qed

To finish the work we will make

{\bf Another proof of theorem \ref{T}:}

First, we study the case of dimension $2$.
Since

\begin{eqnarray*}
A_{0,\vec k}^{2}(t)=-
\int_0^t\f{\dot{\e}_{\vec k}(\tau)}{4{\e}^2_{\vec k}(\tau)}
(i\dot{A}_{1,\vec k}^0-
i\f{\dot{\e}_{\vec k}(t)}{2{\e}_{\vec k}(t)}A_{0,\vec k}^1)d\tau
=-\f{\dot{\e}^2_{\vec k}(t)}{32{\e}^4_{\vec k}(t)}
-\f{1}{2}\left(\int_0^t\f{\dot{\e}^2_{\vec k}(\tau)}
{8{\e}^3_{\vec k}(\tau)}d\tau\right)^2,
\end{eqnarray*}
and  $A_{0,\vec k}^{3}(t)$ is imaginary,
we have

$$|A_{\vec k}(t)|^2=1-\h^2\f{\dot{\e}^2_{\vec k}(t)}
{16{\e}^4_{\vec k}(t)}+h^4 J_{\vec k}(t),$$
with $|J_{\vec k}(t)|\leq \f{\bar{K}}{{\e}^4_{\vec k}}$, where
$\bar{K}$ is a constant independent on $\vec k$
and $\h$.

Starting form this relation, we have 

$$\lim_{\h\rightarrow 0}P^0_{\h}(t)=
\lim_{\h\rightarrow 0}\prod_{\vec k\in\Z^2}|A_{\vec k}(t)|^2=
\lim_{\h\rightarrow 0}\prod_{\vec k\in\Z^2}
\left(1-\h^2\f{\dot{\e}^2_{\vec k}(t)}{16{\e}^4_{\vec k}(t)}\right).$$

We now calcule

\begin{eqnarray*}\prod_{\vec k\in\Z^2}
\left(1-\h^2\f{\dot{\e}^2_{\vec k}(t)}{16{\e}^4_{\vec k}(t)}\right)
=1-\f{\h^2}{1!}\sum_{\vec k\in\Z^2}
\f{\dot{\e}^2_{\vec k}(t)}{16{\e}^4_{\vec k}(t)}+\f{\h^4}{2!}
\sum_{\substack{\vec{k}_1, \vec{k}_2\in \Z^2\\ \vec{k}_1\not=\vec{k}_2}}
\f{\dot{\e}^2_{\vec{k}_1}(t)}{16{\e}^4_{\vec{k}_1}(t)}
\f{\dot{\e}^2_{\vec{k}_2}(t)}{16{\e}^4_{\vec{k}_2}(t)}-\cdots
\end{eqnarray*}

To make this calcul we will use the following

\begin{lemme}
If $n\geq 2$ and $f_{\vec{k}}\geq 0\quad \forall \vec{k}\in\Z^n$, then

\begin{eqnarray*}
\left(\sum_{\vec k}f_{\vec{k}}\right)^n-
\sum_{\substack{\vec{k}_1,\cdots, \vec{k}_n\\ 
\vec{k}_i\not=\vec{k}_j, \mbox{if } i\not=j}}f_{\vec{k}_1}\cdots f_{\vec{k}_n}
\leq \f{n(n-1)}{2}\left(\sum_{\vec k}f_{\vec{k}}\right)^{n-2}
\sum_{\vec k}f^2_{\vec{k}},
\end{eqnarray*}
\end{lemme}
consequently,
\begin{eqnarray*}&&\left|\prod_{\vec k\in\Z^2}
\left(1-\h^2\f{\dot{\e}^2_{\vec k}(t)}{16{\e}^4_{\vec k}(t)}\right)
-\sum_{n=0}^{\infty}\f{1}{n!}
\left(-\h^2\sum_{\vec k\in\Z^2}\f{\dot{\e}^2_{\vec k}(t)}
{16{\e}^4_{\vec k}(t)}\right)^n\right|\\&&\leq
\sum_{n=2}^{\infty}\f{n(n-1)}{2}
\left(\sum_{\vec k\in\Z^2}\f{\h^2\dot{\e}^2_{\vec k}(t)}
{16{\e}^4_{\vec k}(t)}\right)^{n-2}
\sum_{\vec k\in\Z^2}\f{\h^4\dot{\e}^4_{\vec k}(t)}
{16^2{\e}^8_{\vec k}(t)}\f{1}{n!}\\&&\leq
\f{\h^2||\dot{\vec f}||^2_{\infty}}{32}\sum_{n=1}^{\infty}\f{1}{(n-1)!}
\left(\sum_{\vec k\in\Z^2}\f{\h^2\dot{\e}^2_{\vec k}(t)}
{16{\e}^4_{\vec k}(t)}\right)^n.
\end{eqnarray*}

We use that, $\lim_{\h\rightarrow 0}\sum_{\vec k\in\Z^2}
\f{\h^2\dot{\e}^2_{\vec k}(t)}
{16{\e}^4_{\vec k}(t)}=\f{1}{16}\int_{\R^2}
\f{<\dot{\vec f}(t),\vec x>^2}
{(|\vec x|^2+1)^3}d\vec x^2$ and
$\sum_{n=1}^{\infty}\f{x^n}{(n-1)!}=xe^x$, then  
we obtain

\begin{eqnarray*}&&\lim_{\h\rightarrow 0}\left|\prod_{\vec k\in\Z^2}
\left(1-\h^2\f{\dot{\e}^2_{\vec k}(t)}{16{\e}^4_{\vec k}(t)}\right)
-\sum_{n=0}^{\infty}
\left(-\h^2\sum_{\vec k\in\Z^2}\f{\dot{\e}^2_{\vec k}(t)}
{16{\e}^4_{\vec k}(t)}\right)^n\right|
\\&&\leq\lim_{\h\rightarrow 0}\f{\h^2||\dot{\vec f}||^2_{\infty}}{32}
\f{1}{16}\int_{\R^2}
\f{<\dot{\vec f}(t),\vec x>^2}
{(|\vec x|^2+1)^3}d\vec x^2
e^{\f{1}{16}\int_{\R^2}
\f{<\dot{\vec f}(t),\vec x>^2}
{(|\vec x|^2+1)^3}d\vec x^2}
=0.\end{eqnarray*}

By virtue of this result, we have

\begin{eqnarray*}\lim_{\h\rightarrow 0}P^0_{\h}(t)=
\lim_{\h\rightarrow 0}\sum_{n=0}^{\infty}
\f{1}{n!}\left(-\h^2\sum_{\vec k\in\Z^2}\f{\dot{\e}^2_{\vec k}(t)}
{16{\e}^4_{\vec k}(t)}\right)^n=e^{-\f{1}{16}
\int_{\R^2}\f{<\dot{\vec f}(t),\vec x>^2}
{(|\vec x|^2+1)^3}d\vec x^2}. \end{eqnarray*}

Therefore, $\lim_{\h\rightarrow 0}P_{\h}(t)<1$ if 
$\dot{\vec f}(t)\not=\vec 0$.

Now, it is easy to calcule $\lim_{\h\rightarrow 0}P^1_{\h}(t)$.
 In fact, in the first proof of theorem \ref{T}, we have obtained
$$\lim_{\h\rightarrow 0}P^1_{\h}(t)=\lim_{\h\rightarrow 0}
\h^2\sum_{\vec k\in\Z^2}\f{\dot{\e}^2_{\vec k}(t)}
{16{\e}^4_{\vec k}(t)}P^0_{\h}(t),$$
then
$$\lim_{\h\rightarrow 0}P^1_{\h}(t)=
\f{1}{16}
\int_{\R^2}\f{<\dot{\vec f}(t),\vec x>^2}
{(|\vec x|^2+1)^3}d\vec x^2e^{-\f{1}{16}
\int_{\R^2}\f{<\dot{\vec f}(t),\vec x>^2}
{(|\vec x|^2+1)^3}d\vec x^2}.$$

In general, let

$$P^n_{\h}(t)=\f{1}{n!}
\sum_{\substack{\vec{k}_1,\cdots, \vec{k}_n\in \Z^2\\ 
\vec{k}_i\not=\vec{k}_j, \mbox{if } i\not=j}}
|(t)<1^+_{\vec{k}_1}1^-_{-\vec{k}_1}\cdots 
1^+_{\vec{k}_n}1^-_{-\vec{k}_n}|T^t_q|0>(0)|^2,$$
be the probability, that at  $t$ time, does exist $n$ 
particle-antiparticle pairs. Then we have 

$$\lim_{\h\rightarrow 0}P^n_{\h}(t)=
\f{1}{n!}\left(\f{1}{16}
\int_{\R^2}\f{<\dot{\vec f}(t),\vec x>^2}
{(|\vec x|^2+1)^3}d\vec x^2 \right)^n e^{-\f{1}{16}
\int_{\R^2}\f{<\dot{\vec f}(t),\vec x>^2}
{(|\vec x|^2+1)^3}d\vec x^2}.$$

To finish the proof, we have to consider the case of
dimension $3$. We can prove, proceeding as the case of dimension $2$,
 that
$$\lim_{\h\rightarrow 0}P^1_{\h}(t)=\lim_{\h\rightarrow 0}
\h^2\sum_{\vec k\in\Z^3}\f{\dot{\e}^2_{\vec k}(t)}
{16{\e}^4_{\vec k}(t)}P^0_{\h}(t).$$
However,
$\lim_{\h\rightarrow 0}
\h^2\sum_{\vec k\in\Z^3}\f{\dot{\e}^2_{\vec k}(t)}
{16{\e}^4_{\vec k}(t)}=\infty$ if  
$\dot{\vec f}(t)\not=\vec 0$, whence we conclude that,
$\lim_{\h\rightarrow 0}P^0_{\h}(t)=0$ if  
$\dot{\vec f}(t)\not=\vec 0$.

Another proof of last result, is the following

$$\lim_{\h\rightarrow 0}P^0_{\h}(t)=
\lim_{\h\rightarrow 0}\prod_{\vec k\in\Z^3}
\left(1-\h^2\f{\dot{\e}^2_{\vec k}(t)}{16{\e}^4_{\vec k}(t)}\right).$$

Since $\prod_{\vec k\in\Z^3}
\left(1-\h^2\f{\dot{\e}^2_{\vec k}(t)}{16{\e}^4_{\vec k}(t)}\right)\leq
\prod_{\vec k\in\Z^3}
\left(1-\f{L\h^3\dot{\e}^2_{\vec k}(t)}{16{\e}^4_{\vec k}(t)}\right)$
if $L\h\leq 1$, we obtain

\begin{eqnarray*}
&&\lim_{\h\rightarrow 0}P^0_{\h}(t)\leq
\lim_{L\rightarrow\infty}\lim_{\h\rightarrow 0}\prod_{\vec k\in\Z^3}
\left(1-\f{L\h^3\dot{\e}^2_{\vec k}(t)}{16{\e}^4_{\vec k}(t)}\right)
=\lim_{L\rightarrow\infty}
e^{-\f{L}{16}
\int_{\R^3}\f{<\dot{\vec f}(t),\vec x>^2}
{(|\vec x|^2+1)^3}d\vec x^2}\\&&
=\left\{\begin{array}{ccc}
0&\mbox{if}& \dot{\vec f}(t)\not=\vec 0\\
1&\mbox{if}& \dot{\vec f}(t)=\vec 0.
\end{array}\right.\end{eqnarray*}

\qed

\section{References}

\begin{enumerate}

\item[{[1]}] {\em J.D.BJORKEN and S.D.DRELL, Relativistic Quantum Fields;
McGraw-Hill Book Co., New York (1965).}

\item[{[2]}] {\em P.A.M.DIRAC, The principles of quantum mechanics;
Oxford University Press (1958).}

\item[{[3]}] {\em S.A.FULLING, Aspects of Quantum Field Theory in
Curved Spacetime; Cambridge University Press (1989).}

\item[{[4]}] {\em W.GREINER, B.M\"ULLER, J.RAFELSKI, Quantum
Electrodynamics of Strong Fields; Springer-Verlang (1985).}

\item[{[5]}] {\em G.A.HAGEDORN, Semiclassical quantum mechanics I.
The $\hbar\rightarrow 0$ limit for coherent states; Comm. Math.
Phys. 71, no. 1, pag. 77-93, (1980).}

\item[{[6]}] {\em J.HARO, El l{\'\i}mit cl\`assic de la mec\`anica
qu\`antica; Tesi Doctoral, U.A.B. (1997).}

\item[{[7]}] {\em J.HARO, \'Etude classique de l'\'equation de Dirac;
Ann. Fond. Louis de Broglie 23, no. 3-4, pag. 166-172, (1998).}

\item[{[8]}] {\em J.HARTHONG, \'Etudes sur la m\'ecanique quantique;
Asterisque, 111, (1984).}

\item[{[9]}] {\em V.P.MASLOV and M.V.FEDORIUK, Semi-classical 
aproximation
in quantum mechanics; D.Riedel Publishing Compay, Dordrecht, Holland
(1981).}

\end{enumerate}

\end{document}